\documentclass[12pt,a4paper]{article}
\usepackage{amssymb,amsmath}
\usepackage{amsfonts}
\usepackage{enumerate}
\usepackage{a4}
\newtheorem{defi}{Definition}
\newtheorem{prop}{Proposition}
\newtheorem{theorem}{Theorem}
\newtheorem{lemma}{Lemma}
\newtheorem{cor}{Corollary}

\newtheorem{eg}{Example}
\begin{document}
 \baselineskip15pt
	
	\title{\bf Two Dimensional $\left( \alpha,\beta \right) $-Constacyclic Codes of arbitrary length  over a Finite Field }
	\author{Swati Bhardwaj ~ and ~ Madhu Raka{\footnote {The research  is supported by CSIR, sanction no. ES/ 21(1042)/17/ EMR-II.}}
		\\ \small{\em Centre for Advanced Study in Mathematics}\\
		\small{\em Panjab University, Chandigarh-160014, INDIA}\\ \small emails: swatibhardwaj2296@gmail.com, mraka@pu.ac.in\\
		\date{}}
	\maketitle
\maketitle
\begin{abstract}  In this paper we  characterize the algebraic structure of  two-dimensional $(\alpha,\beta )$-constacyclic codes of arbitrary length $s.\ell$ and of their duals. For $\alpha,\beta \in \{1,-1\}$, we give necessary and sufficient conditions for a two-dimensional $(\alpha,\beta )$-constacyclic code to be self-dual. We also show that a two-dimensional $(\alpha,1 )$-constacyclic code $\mathcal{C}$ of length $n=s.\ell$  can not be self-dual  if $\gcd(s,q)= 1$. Finally, we give some examples of self-dual, isodual,  MDS and quasi-twisted codes corresponding to two-dimensional $(\alpha,\beta )$-constacyclic codes.\vspace{2mm}
		\end{abstract}
	
	\section{ Introduction}
	
One of the important generalizations of the cyclic code is two-dimensional cyclic
 code. Let $\mathbb{F}_q$ be the finite field with $q=p^m$ elements. Suppose that $C$  is a linear code over $\mathbb{F}_q$ of length $s.\ell$ whose codewords
are viewed as $s\ell$ arrays. That is every codeword $c$ in $C$ has the following form
$$c =\left( \begin{array}{cccc}
       c_{0,0} & c_{0,1} & ... & c_{0,\ell-1}  \\
       c_{1,0} & c_{1,1} & ... & c_{1,\ell-1}  \\
       ... & ... & ... & ...  \\
        c_{s-1,0} &c_{s-1,1} & ... & c_{s-1,\ell-1}
     \end{array}\right).$$

\noindent If $C$ is closed under row shift and column shift of codewords, then we call $C$ a two-dimensional
cyclic code of length $s.\ell$ over $\mathbb{F}_q$. \vspace{2mm}

In 1975, Ikai et al. \cite{Ikai} introduced the concept of two-dimensional  codes and in 1977, Imai \cite{Imai} characterized binary two-dimensional cyclic codes.  G$\ddot{\rm u}$neri and $\ddot{\rm O}$zbudak \cite{Gun2012} studied the relations between quasi-cyclic codes and two-dimensional cyclic codes.	In 2014, Xiuli et al. \cite{Xi} generalized this concept to skew cyclic two-dimensional codes over a finite field. It is well known that a two-dimensional  cyclic
code of length $s.\ell$ over $\mathbb{F}_q$ is an ideal
of the polynomial ring $\mathbb{F}_q[x, y]/\langle x^s - 1, y^{\ell} -1 \rangle$. In 2016,  Sepasdar and  Khashyarmanesh \cite{Sepas2016} characterized the structure of
two-dimensional cyclic codes of length $s.2^k$ over $\mathbb{F}_{p^m} $ iteratively,  where $p$ is an odd prime. In \cite{Sepas2017},  Sepasdar  characterized the generator matrix of two-dimensional cyclic codes of  length $s.\ell$.\vspace{2mm}

Constacyclic codes over finite fields have a very significant role in the theory of error-correcting codes. A lot of work on constacyclic codes has been done in recent years, see for example \cite{BR, GR5, R, RKG}. Given nonzero elements $\alpha$ and $\beta$  of  $\mathbb{F}_q $, a two-dimensional $(\alpha,\beta )$-constacyclic code of length $s.\ell$ is an ideal of the polynomial ring $\mathbb{F}_q[x, y]/ <x^s-\alpha, y^{\ell}-\beta >$.  In 2018, Rajabi and Khashyarmanesh \cite {RajKhas}  investigated some repeated-root two-dimensional $(\alpha,\beta )$-constacyclic codes of length $2p^s.2^k$ over $\mathbb{F}_{p^m} $, where $p$ is an odd prime, using the structure of \cite{Sepas2016}. They also studied  self-dual repeated-root two-dimensional $(\alpha, 1)$-constacyclic codes of length $2p^s.2$ and $2p^s.2^2$ over $\mathbb{F}_{p^m} $ for $\alpha \in \{-1, 1\}$. \vspace{2mm}

In this paper we give a novel method to characterize the algebraic structure of  two-dimensional $(\alpha,\beta )$-constacyclic codes of arbitrary length $s.\ell$ and of their duals
over a finite field $\mathbb{F}_q $, using central primitive idempotents of the ring $\mathbb{F}_q[ y]/ < y^{\ell}-\beta >$ (Theorems 1-3). Our method is quite different from that of \cite{RajKhas, Sepas2016, Sepas2017} and it does not require $s$ to be a multiple of $p$, the characteristic of $\mathbb{F}_q $ as in \cite {RajKhas}.  When $\alpha,\beta \in \{1,-1\}$, we give necessary and sufficient conditions for a two-dimensional $(\alpha,\beta )$-constacyclic code to be self-dual (Theorem \ref{th4}). We also prove that a two-dimensional $(\alpha,1 )$-constacyclic code $\mathcal{C}$ of length $n=s.\ell$  can not be self-dual   if $\gcd(s,q)= 1$, for any $s$, where $\alpha \in \{1,-1\}$ (Theorem \ref{th5}). As an illustration, we give some examples of  self-dual or isodual,  MDS or near MDS and quasi-twisted codes in Section 4.
\section{Preliminaries}
Let $\alpha$ and $\beta$ be two non-zero elements of field $\mathbb{F}_q $. For a two-dimensional codeword
$$c =\left( \begin{array}{cccc}
       c_{0,0} & c_{0,1} & ... & c_{0,\ell-1}  \\
       c_{1,0} & c_{1,1} & ... & c_{1,\ell-1}  \\
       ... & ... & ... & ...  \\
        c_{s-1,0} &c_{s-1,1} & ... & c_{s-1,\ell-1}
     \end{array}\right)$$
 its row $\alpha$-constacyclic shift and column $\beta$-constacyclic shift are  defined as

 $$\sigma_{\alpha}(c) =\left( \begin{array}{cccc}
       \alpha c_{s-1,0} &\alpha c_{s-1,1} & ... & \alpha c_{s-1,\ell-1}  \\
       c_{0,0} & c_{0,1} & ... & c_{0,\ell-1}  \\
       ... & ... & ... & ...  \\
        c_{s-2,0} &c_{s-2,1} & ... & c_{s-2,\ell-1}
     \end{array}\right) {\rm ~~~~ and}$$

$$\tau_{\beta}(c) =\left( \begin{array}{cccc}
       \beta c_{0,\ell-1} & c_{0,1} & ... & c_{0,\ell-2}  \\
       \beta c_{1,\ell-1} & c_{1,1} & ... & c_{1,\ell-2}  \\
       ... & ... & ... & ...  \\
       \beta c_{s-1,\ell-1} &c_{s-1,1} & ... & c_{s-1,\ell-2}
     \end{array}\right).$$
\begin{defi}\normalfont If a linear two-dimensional code $\mathcal{C}$ is closed under both row $\alpha$-constacyclic shift $\sigma_{\alpha}$ and column $\beta$-constacyclic shift $\tau_{\beta}$, then it is called a two-dimensional $\left( \alpha,\beta \right) $-constacyclic code of  length $s.\ell$ over $\mathbb{F}_q $.\end{defi}

Each codeword $c$ in a two-dimensional $(\alpha,\beta )$-constacyclic code has a unique polynomial representation
$$c(x,y)=f_0(x)+f_1(x)y+\cdots +f_{\ell-1}(x)y^{\ell-1}, {\rm ~where~} f_i(x)=c_{0,i}+c_{1,i}x+\cdots+c_{s-1,i}x^{s-1} $$
 for $i=0,1,\cdots,\ell-1$. $xc(x,y)$ corresponds to the codeword $\sigma_{\alpha}(c)$ and $yc(x,y)$ corresponds to $\tau_{\beta}(c)$. Therefore a two-dimensional $(\alpha,\beta )$-constacyclic code can be regarded as an ideal in the polynomial ring $\mathbb{F}_q[x, y]/ <x^s-\alpha, y^{\ell}-\beta >$.

 \begin{defi}\normalfont Let \vspace{2mm}

$\begin{array}{ll}a&= \big(a_{0,0},a_{0,1},\cdots,a_{0,\ell-1}|a_{1,0},\cdots,a_{1,\ell -1}|\cdots|a_{s-1,0},\cdots,a_{s-1,\ell-1}\big)\vspace{2mm}\\& = \big(a^{(0)}|a^{(1)}|\cdots |a^{(s-1)}\big)\end{array}$ \vspace{1mm}

\noindent be a vector in  $\mathbb{F}_q^{n}$ divided into $s$ equal parts of length $\ell$ where $n=s\ell$. A linear code $\mathcal{C}$ is called a $\lambda$- quasi-twisted code of index $s$ if $\big(\lambda a^{(s-1)}|a^{(0)}|\cdots |a^{(s-2)}\big)\in \mathcal{C}$ whenever $a \in \mathcal{C}$.  When $\lambda=1$, $\mathcal{C}$ is called quasi-cyclic code of index $s$. \end{defi}
For a two-dimensional code  $\mathcal{C}$, let
\begin{equation}\label{eq1}\mathcal{C}_1= \left\{  \big(c_{0,0},\cdots,c_{0,\ell-1}|c_{1,0},\cdots,c_{1,\ell -1}|\cdots|c_{s-1,0},\cdots,c_{s-1,\ell-1}\big) :c \in \mathcal{C} \right \} \end{equation}
 \begin{equation}\label{eq2}\mathcal{C}_2= \left\{  \big(c_{0,0},\cdots,c_{s-1,0}|c_{0,1},\cdots,c_{s-1, 1}|\cdots|c_{0,\ell-1},\cdots,c_{s-1,\ell-1}\big) :c \in \mathcal{C} \right \}. \end{equation}
Clearly $\mathcal{C}_1$ and $\mathcal{C}_2$ are permutation equivalent. By definition, $\mathcal{C}$ is a two-dimensional $(\alpha,\beta )$-constacyclic code if and only if $\mathcal{C}_1$ is an $\alpha$- quasi-twisted code of index $s$ and $\mathcal{C}_2$ is  a $\beta$- quasi-twisted code of index $\ell$.\vspace{2mm}

 \noindent For a linear code $\mathcal{C}$ of length $n$ over $\mathbb{F}_q$, the dual code  $\mathcal{C}^\bot$ is defined as $\mathcal{C}^\bot =\{x\in \mathbb{F}_q^n~ |~ x \cdot y=0 ~ {\rm for ~ all~} y \in \mathcal{C}\}$, where $x\cdot y$ denotes the usual Euclidean inner product. If $\mathcal{C}$ is $\lambda$-constacyclic code over $\mathbb{F}_q$, then $\mathcal{C}^\bot$ is a $\lambda^{-1}$-constacyclic code over $\mathbb{F}_q$. A similar result holds for a  two-dimensional $(\alpha,\beta )$-constacyclic code. Following results are due to \cite{RajKhas}.

\begin{prop} Let $f(x,y)=f_0(x)+f_1(x)y+\cdots +f_{\ell-1}(x)y^{\ell-1}, g(x,y)=g_0(x)+g_1(x)y+\cdots +g_{\ell-1}(x)y^{\ell-1} \in \mathbb{F}_q[x, y]$ , where $f_i(x)=a_{0,i}+a_{1,i}x+\cdots+a_{s-1,i}x^{s-1} $ and $g_i(x)=b_{0,i}+b_{1,i}x+\cdots+b_{s-1,i}x^{s-1} $ for $i=0,1,\cdots, \ell-1$. Then
$$f(x,y)g(x,y)=0 {\rm ~~in~~} \mathbb{F}_q[x, y]/ <x^s-\alpha, y^{\ell}-\beta >$$
if and only if $(a_0,a_1,\cdots, a_{\ell-1})$ is orthogonal to $(b_{\ell-1},b_{\ell-2},\cdots, b_0)$ and all its $(\alpha^{-1},\beta^{-1} )$-constacyclic shifts where $a_i=(a_{0,i},a_{1,i},\cdots,a_{s-1,i})$ and \\ $b_i=(b_{0,i},b_{1,i},\cdots,b_{s-1,i})$. \end{prop}

\begin{prop} The dual of a two-dimensional $(\alpha,\beta )$-constacyclic code is a two-dimensional $(\alpha^{-1},\beta^{-1} )$-constacyclic code.\end{prop}

Let $S$ be a non-empty subset of a commutative ring $R$. The annihilator of $S$, denoted by ann$(S)$, is the set ann$(S)= \{f : fg=0 {\rm ~for~ all~} g \in S\}$. Clearly ann$(S)$ is an ideal of $R$. For a polynomial $f(x)$ with $\deg(f(x))=k$, its reciprocal is defined as $f^\ast(x)=x^kf(1/x)$ and $S^\ast=\{ f^\ast : f\in S\}$. If $C$ is a $\lambda$-constacyclic code of length $n$ over $\mathbb{F}_q$ generated by $g(x)$, then
$C^\perp$ is $\lambda^{-1}$-constacyclic code generated by 	$h^*(x)$ where $x^n-\lambda= g(x)h(x)$.
\begin{prop} Suppose that $\alpha,\beta \in \{1,-1\}$. Let $\mathcal{C}$ be a two-dimensional $(\alpha,\beta )$-constacyclic code, then $\mathcal{C}^\perp$ is also a two-dimensional $(\alpha,\beta )$-constacyclic code and $\mathcal{C}^\perp = ({\rm ann}(\mathcal{C}))^\ast$, also denoted as   ann$^\ast(\mathcal{C})$. \end{prop}

\section{Two Dimensional $\left( \alpha,\beta \right) $-Constacyclic Codes of  length $s.\ell$}
In this section we will obtain generators of a two-dimensional $(\alpha,\beta )$-constacyclic code of arbitrary length $s.\ell$ and that of its dual. First we study primitive central idempotents of the ring $\mathbb{F}_q[y]/\langle y^\ell-\beta\rangle $ and discuss some of their properties.\vspace{2mm}

Let  $\mathcal{R}$ denote the polynomial ring $\mathbb{F}_q[x,y]/\langle x^s-\alpha,y^{\ell}-\beta \rangle$ where $\alpha,\beta \in \mathbb{F}_q^*$. Let $r$ be the order of $\beta$ in $\mathbb{F}_q$ so that $\beta^r=1$ and $r|(q-1)$. Let $\omega$ be an $r\ell^{th}$ root of unity such that $\omega^{\ell}=\beta$. Assume that $q \equiv 1 \pmod {r\ell}$  so that $\omega \in \mathbb{F}_q$, since for some integer $k$, $\omega^{q-1}=\omega^{r\ell k}=\beta^{rk}=1$.  Then,
\begin{equation*}
y^{r\ell}-1=(y-1)(y-\omega)(y-\omega^2) \cdots (y-\omega^{r\ell-1}).
\end{equation*}
\noindent Define
\begin{equation*}\begin{array}{ll}
\zeta_0(y) &=\displaystyle \frac{(y-\omega)(y-\omega^2)(y-\omega^3)\cdots(y-\omega^{r\ell-1})}{(1-\omega)(1-\omega^2)(1-\omega^3)\cdots
(1-\omega^{r\ell-1})}, \vspace{2mm}\\
\zeta_1(y) &=\displaystyle \frac{(y-1)(y-\omega^2)(y-\omega^3)\cdots(y-\omega^{r\ell-1})}{(\omega-1)(\omega-\omega^2)(\omega-\omega^3)
\cdots(\omega-\omega^{r\ell-1})}, \vspace{2mm}\\
&\vdots \end{array}\end{equation*}

\begin{equation*}\begin{array}{ll}
\zeta_j(y) &=\displaystyle \frac{(y-1)(y-\omega)\cdots(y-\omega^{j-1})(y-\omega^{j+1})\cdots(y-\omega^{r\ell-1})}{(\omega^j-1)
(\omega^j-\omega)\cdots(\omega^j-\omega^{j-1})(\omega^j-\omega^{j+1})\cdots(\omega^j-\omega^{r\ell-1})},\vspace{2mm}\\
&\vdots \vspace{2mm}\\
\zeta_{r\ell-1}(y) &=\displaystyle \frac{(y-1)(y-\omega)(y-\omega^2)\cdots(y-\omega^{r\ell-2})}{(\omega^{r\ell-1}-1)(\omega^{r\ell-1}-\omega)
(\omega^{r\ell-1}-\omega^2)\cdots(\omega^{r\ell-1}-\omega^{r\ell-2})},
\end{array}\end{equation*}

\noindent then  $\zeta_0(y),\zeta_1(y),\cdots,\zeta_{r\ell-1}(y)$ are primitive central idempotents in $\mathbb{F}_q[y]/\langle y^{r\ell}-1\rangle $ $i.e.$ $\zeta_0(y)+\zeta_1(y)+ \cdots +\zeta_{r\ell-1}(y)=1$ and $\zeta_i(y) \zeta_j(y)= \delta_{i,j} \zeta_i(y)$ for $i,j \in \{0,1,2, \cdots ,r\ell-1 \}$, where $\delta_{i,j}$ is the Kronecker Delta function. For a proof of it see \cite{GR2}. Also, $\zeta_j(\omega^j)=1$ and $\zeta_j(\omega^k)=0$ for $j \neq k$ in the ring $\mathcal{R}$.

\begin{lemma} \label{lem1} \normalfont
	For $j=0,1,2,\cdots,r\ell-1$, we have $$\zeta_j(y)=\displaystyle \frac{1}{r\ell} \Big( 1+\omega^{r\ell-j}y+(\omega^{r\ell-j}y)^2+\cdots+(\omega^{r\ell-j}y)^{r\ell-1} \Big).$$
\end{lemma}

\noindent {\bf Proof :}
For any $j$,  $j=0,1,\cdots,r\ell-1$, the numerator of  $\zeta_j(y)$ is
\begin{equation*} \begin{array}{l}
(y-1)\cdots(y-\omega^{j-1})(y-\omega^{j+1})\cdots(y-\omega^{r\ell-1})\vspace{2mm} \\\hspace{4mm}=\displaystyle \frac{y^{r\ell}-1}{y-\omega^j} \vspace{2mm}\\
\hspace{4mm}=\displaystyle \frac{y^{r\ell}-(\omega^j)^{r\ell}}{y-\omega^j} ~~~~~(\because \omega^{r\ell}=1) \vspace{2mm}\\
\hspace{4mm}= y^{r\ell-1}+\omega^j y^{r\ell-2}+(\omega^j)^2 y^{r\ell-3}+\cdots
 +(\omega^j)^{r\ell-2} y+(\omega^j)^{r\ell-1} \vspace{2mm}\\
\hspace{4mm}=(\omega^j)^{r\ell-1}\big\{1+(\omega^j)^{-r\ell+1}(\omega^j)^{r\ell-2}y +
(\omega^j)^{-r\ell+1}(\omega^j)^{r\ell-3}y^2+ \cdots \vspace{1mm}\\
~~~~~~~~~~ +(\omega^j)^{-r\ell+1}\omega^j y^{r\ell-2}+(\omega^j)^{-r\ell+1}y^{r\ell-1} \big\} \vspace{2mm}\\
\hspace{4mm}= \omega^{-j}\big\{1+(\omega^j)^{-1}y+(\omega^j)^{-2}y^2+\cdots
~~+(\omega^j)^{-r\ell+2}y^{r\ell-2}+(\omega^j)^{-r\ell+1}y^{r\ell-1} \big\} \vspace{2mm}\\
\hspace{4mm}= \displaystyle \frac{1}{\omega^j}\left\{ 1+\omega^{r\ell-j}y+(\omega^{r\ell-j}y)^2+\cdots
~~+(\omega^{r\ell-j}y)^{r\ell-1} \right\}.
\end{array} \end{equation*}

\noindent To compute denominator of $\zeta_j(y)$,  substitute $y=\omega^j$ in the above expression and get

\begin{equation*} \begin{array}{l}(\omega^j-1)\cdots(\omega^j-\omega^{j-1})(\omega^j-\omega^{j+1})\cdots(\omega^j-\omega^{r\ell-1})\vspace{2mm} \\\hspace{4mm}=\displaystyle \frac{1}{\omega^j}\big\{ 1+\omega^{r\ell-j}\omega^j+\cdots+(\omega^{r\ell-j}\omega^j)^{r\ell-1} \big\} \vspace{2mm} \\\hspace{4mm}=\displaystyle \frac{1}{\omega^j}\{1+1+\cdots+1\} \vspace{2mm} =\displaystyle \frac{r\ell}{\omega^j}.
\end{array}\end{equation*}

\noindent This gives the desired result.
\hfill $\square$

\begin{lemma} \label{lem2} \normalfont
	$\zeta_{1+kr}(y) y^j = (\omega^{1+kr})^j \zeta_{1+kr}(y)$ for $~~j,k \in \{ 0,1,2, \cdots ,\ell-1\} $.
\end{lemma}

\noindent {\bf Proof :}
In the ring $\mathbb{F}_q[y]/\langle y^{r\ell}-1 \rangle$, using Lemma \ref{lem1}
\begin{equation*}\begin{array}{ll}
\zeta_{1+kr}(y) y &= \displaystyle \frac{1}{r\ell}  \Big( 1+\omega^{r\ell-1-kr}y+(\omega^{r\ell-1-kr}y)^2+\cdots+(\omega^{r\ell-1-kr}y)^{r\ell-1} \Big)y \vspace{2mm}\\
&= \displaystyle \frac{1}{r\ell} \Big( y+\omega^{r\ell-1-kr}y^2+\omega^{2(r\ell-1-kr)}y^3+\cdots+\omega^{(r\ell-1)(r\ell-1-kr)}y^{r\ell} \Big) \vspace{2mm}\\
&= \displaystyle \frac{1}{r\ell} \Big( \omega^{1+kr}+y+\omega^{r\ell-1-kr}y^2+\omega^{2(r\ell-1-kr)}y^3+\cdots+\omega^{(r\ell-2)(r\ell-1-kr)}y^{r\ell-1} \Big) \vspace{2mm}\\
&= \displaystyle \frac{1}{r\ell}\omega^{1+kr} \Big( 1+\omega^{r\ell-1-kr}y+(\omega^{r\ell-1-kr}y)^2+\cdots+(\omega^{r\ell-1-kr}y)^{r\ell-1} \Big) \vspace{2mm}\\
&= \omega^{1+kr}\zeta_{1+kr}(y). \vspace{2mm}\\
\zeta_{1+kr}(y) y^2 &= \Big(\zeta_{1+kr}(y) y\Big) y \vspace{2mm}\\
&= \omega^{1+kr}\zeta_{1+kr}(y) y \vspace{2mm}\\
&=(\omega^{1+kr})^2\zeta_{1+kr}(y) \vspace{2mm}\\
& \vdots \vspace{2mm}\\
\zeta_{1+kr}(y) y^j &= (\omega^{1+kr})^j \zeta_{1+kr}(y).~~~~~~~~~~~~~~~~~~~~~~~~~~~~~~~~~~~~~~~~~~~~~~~~~~~~~~~~ \Box
\end{array}\end{equation*}\vspace{2mm}

\noindent  Following the notations of \cite{GR5}, let $P_{\ell,\beta}=\{1+rk : k=0,1,2,\cdots,\ell-1 \}$. Then
\begin{equation}\begin{array}{ll}
y^\ell-\beta&=(y-\omega)(y-\omega^{1+r})(y-\omega^{1+2r})\cdots(y-\omega^{1+(\ell-1)r})\vspace{2mm}\\
&=\displaystyle\prod_{i \in P_{\ell,\beta}} \left(y-\omega^i\right)
\end{array}\end{equation}

\noindent Let $y^{r\ell}-1=(y^\ell-\beta)Q(y)$ where $Q(y)= \displaystyle\prod_{i \notin P_{\ell,\beta}} \left(y-\omega^i\right)$.
 Define
\begin{equation*}\begin{array}{ll}
\eta_0(y)&=\displaystyle \frac{(y-\omega^{1+r})(y-\omega^{1+2r})\cdots(y-\omega^{1+(\ell-1)r})}{(\omega-\omega^{1+r})(\omega-\omega^{1+2r})
\cdots(\omega-\omega^{1+(\ell-1)r})}, \vspace{2mm}\\
\eta_{1}(y)&=\displaystyle \frac{(y-\omega)(y-\omega^{1+2r})\cdots(y-\omega^{1+(\ell-1)r})}{(\omega^{1+r}-\omega)(\omega^{1+r}-\omega^{1+2r})
\cdots(\omega^{1+r}-\omega^{1+(\ell-1)r})}, \vspace{2mm}\\
&\vdots \vspace{2mm}\\
\eta_{\ell-1}(y)&=\displaystyle \frac{(y-\omega)(y-\omega^{1+r})\cdots(y-\omega^{1+(\ell-2)r})}{(\omega^{1+(\ell-1)r}-\omega)(\omega^{1+(\ell-1)r}
-\omega^{1+r})\cdots(\omega^{1+(\ell-1)r}-\omega^{1+(\ell-2)r})},
\end{array}\end{equation*}

\noindent then  $\eta_0(y),\eta_{1}(y),\cdots,\eta_{\ell-1}(y)$ are primitive central idempotents in $\mathbb{F}_q[y]/\langle y^\ell-\beta\rangle $ $i.e.$ $\eta_0(y)+\eta_{1}(y)+ \cdots +\eta_{\ell-1}(y)=1$ and $\eta_i(y) \eta_j(y)= \delta_{i,j} \eta_i(y)$ for $i,j \in \{0,1, \cdots ,\ell-1 \}$. \\ Note that $\eta_{j}(y)=\zeta_{j+1}(y)$ and  $\eta_{\ell-1}(y)=\zeta_0(y)$ if $r=1$.

\begin{lemma} \label{lem3}\normalfont
	For $k=0,1,2,\cdots,\ell-1$ we have $\zeta_{1+kr}(y)=\eta_{k}(y) \left(  \frac{Q(y)}{c_{k}} \right) $ for some constant $c_{k}$ in $\mathbb{F}_q^*$.
\end{lemma}

\noindent {\bf Proof :}
\begin{equation*}\begin{array}{ll}
\zeta_{1+kr}(y)&= \displaystyle \frac{(y-1)(y-\omega) \cdots (y-\omega^{kr})(y-\omega^{kr+2}) \cdots (y-\omega^{r\ell-1})}{(\omega^{1+kr}-1)(\omega^{1+kr}-\omega) \cdots (\omega^{1+kr}-\omega^{kr})(\omega^{1+kr}-\omega^{kr+2}) \cdots (\omega^{1+kr}-\omega^{r\ell-1})} \vspace{2mm}\\
&= \displaystyle \frac{ (y-\omega)(y-\omega^{1+r}) \cdots(y-\omega^{1+(k-1)r})(y-\omega^{1+(k+1)r}) \cdots (y-\omega^{1+(\ell-1)r})  \cdot Q(y) }{(\omega^{1+kr}-1)(\omega^{1+kr}-\omega) \cdots (\omega^{1+kr}-\omega^{kr})(\omega^{1+kr}-\omega^{kr+2}) \cdots (\omega^{1+kr}-\omega^{r\ell-1})} \vspace{2mm}\\
&= \eta_{k}(y)   \frac{Q(y)}{c_{k}}.~~~~~~~~~~~~~~~~~~~~~~~~~~~~~~~~~~~~~~~~~~~~~~~~~~~~~~~~~~~\Box
\end{array}\end{equation*}

\begin{lemma}\label{lem4} \normalfont
	$\eta_{k}(y) y^j = \big(\omega^{1+kr}\big)^j \eta_{k}(y)$ for $j,k \in \{ 0,1,2, \cdots ,\ell-1\} $.
\end{lemma}

\noindent {\bf Proof :}
Using Lemmas \ref{lem2} and  \ref{lem3}, we see that
\begin{equation*}\begin{array}{ll}
\eta_{k}(y) y^j &= \zeta_{1+kr}(y) y^j \displaystyle \frac{c_{k}}{Q(y)} \vspace{2mm}\\
&= (\omega^{1+kr})^j \zeta_{1+kr}(y) \displaystyle \frac{c_{k}}{Q(y)}  \vspace{2mm}\\
&= (\omega^{1+kr})^j\eta_{k}(y)~~~~~~~~~~~~~~~~~~~~~~~~~~~~~~~~~~~~~~~~~~~~~~ \Box
\end{array} \end{equation*}
\begin{lemma} \label{lem5} \normalfont The reciprocal polynomials of $\eta_k(y)$, for $k=0,1,\cdots,\ell-1,$ are given by
\begin{equation*} \eta_k^\ast(y) = \left\{ \begin{array}{lll}b_k ~\eta_{\ell-2-k}(y) & {\rm if} & \beta=1 \\
b_k~ \eta_{\ell-1-k}(y) & {\rm if} & \beta=-1 \end{array}\right.\end{equation*}
for some constant $b_k \in \mathbb{F}_q^\ast$.
\end{lemma}
\noindent {\bf Proof :} When $\beta=1$, we have $r=1$. Then by definition $\eta_k(y)=\zeta_{k+1}(y)$.
As $\zeta_{k+1}(y)= \frac{1}{a_{k}}\frac{y^{\ell}-1}{y-\omega^{k+1}}$, for some constant $a_k$ and $\omega^{\ell}=1$, we have  $$\zeta_{k+1}^\ast(y)=\frac{1}{a_{k}\omega^{k+1}}\frac{y^{\ell}-1}{(y-\omega^{-(k+1)})}=b_k ~\zeta_{\ell-k-1}(y)= b_k ~\eta_{\ell-2-k}(y)$$ for some constant $b_k$.\vspace{2mm}

When  $\beta=-1$, we have $r=2$. As $\eta_{k}(y)= \frac{1}{a_{k}}\cdot\frac{y^{\ell}+1}{y-\omega^{1+2k}}$, for some constant $a_k$, we get
$$\eta_{k}^\ast(y)=\frac{1}{a_{k}~\omega^{1+2k}}\frac{y^{\ell}+1}{(y-\omega^{-(1+2k)})}=b_k~\frac{y^{\ell}+1}
{y-\omega^{1+2(\ell-1-k)}}=b_k ~\eta_{\ell-1-k}(y)$$ for some constant $b_k$, since $\omega^{2\ell}=1$. \hfill $\Box$

\subsection{Generator matrix}

\noindent Let $\mathcal{C}$ be a two dimensional $(\alpha,\beta )$-constacyclic code i.e. $\mathcal{C}$ is an ideal of the ring $\mathcal{R}=\mathbb{F}_q[x,y]/\langle x^s-\alpha,y^\ell-\beta \rangle$.
 Define,

\begin{equation*} \begin{array}{ll}
I_0 &= \{ f(x) \in \mathbb{F}_q[x]/\langle x^s-\alpha \rangle : \eta_0(y)f(x) \in \mathcal{C} \} \vspace{2mm}\\
I_1 &= \{ f(x) \in \mathbb{F}_q[x]/\langle x^s-\alpha \rangle : \eta_{1}(y)f(x) \in \mathcal{C} \} \vspace{2mm}\\
I_2 &= \{ f(x) \in \mathbb{F}_q[x]/\langle x^s-\alpha \rangle : \eta_{2}(y)f(x) \in \mathcal{C} \} \vspace{2mm}\\
&\vdots \vspace{2mm}\\
I_{\ell-1} &= \{ f(x) \in \mathbb{F}_q[x]/\langle x^s-\alpha \rangle : \eta_{\ell-1}(y)f(x) \in \mathcal{C} \}.
\end{array}\end{equation*}

\noindent Then $I_0,I_1,\cdots,I_{l-1}$ are ideals in the principal ideal ring $\mathbb{F}_q[x]/\langle x^s-\alpha \rangle$ and are therefore principal.  Hence there exist unique monic generator polynomials $p_j(x)$, such that $I_j=\langle p_j(x) \rangle$ where $p_j(x)|x^s-\alpha$ for $j=0,1,\cdots,\ell-1$.
These $I_j$'s are  $\alpha$ - constacyclic codes of length $s$. The polynomials $p_j(x)$ in $\mathbb{F}_q[x]/\langle x^s-\alpha \rangle$ and $\eta_{j}(y)$ in $\mathbb{F}_q[y]/\langle y^\ell-\beta \rangle$ contribute to the generators of $\mathcal{C}$.

\begin{theorem} \label{th1} Let $\mathcal{C}$ be an ideal in the ring $\mathcal{R}=\mathbb{F}_q[x,y]/\langle x^s-\alpha,y^\ell-\beta \rangle$, then
\begin{equation}\label{eq4}\mathcal{C} = \big\langle \eta_0(y)p_0(x), \eta_{1}(y)p_1(x), \cdots , \eta_{\ell-1}(y)p_{\ell-1}(x) \big\rangle. \end{equation}
\end{theorem}

\noindent {\bf Proof :}
Let $g(x,y)$ be an arbitrary element of $\mathcal{C}$, therefore there exist polynomials $g_i(x) \in \mathbb{F}_q[x]/\langle x^s-\alpha \rangle$ for $i=0,1,\cdots,\ell-1$ such that $$g(x,y)=g_0(x)+g_1(x)y+\cdots+g_{\ell-1}(x)y^{\ell-1}.$$
Then for each $k,0\le k \le \ell-1$, using Lemma 4, we have
\begin{equation} \begin{array}{ll} \label{eq3}
g(x,y)\eta_{k}(y)&=g_0(x)\eta_{k}(y)+g_1(x)\eta_{k}(y) y+\cdots+g_{\ell-1}(x)\eta_{k}(y)
y^{\ell-1} \vspace{2mm}\\
&=g_0(x)\eta_{k}(y)+g_1(x)\omega^{1+kr}\eta_{k}(y)+\cdots+g_{\ell-1}(x)(\omega^{1+kr})^{\ell-1}
\eta_{k}(y) \vspace{2mm}\\
&=\eta_{k}(y)\big(g_0(x)+g_1(x)\omega^{1+kr}+\cdots+g_{\ell-1}(x)(\omega^{1+kr})^{\ell-1}\big)\vspace{2mm}\\
&=\eta_{k}(y)g(x,\omega^{1+kr}).
\end{array} \end{equation}

\noindent Now, $g(x,y) \in \mathcal{C}$ implies $g(x,y)\eta_{k}(y) \in \mathcal{C}$, as $\mathcal{C}$ is an ideal and hence by definition of $I_k$,  $g(x,\omega^{1+kr}) \in I_k=\langle p_k(x) \rangle. $ Thus, there exist some $ p_k''(x) \in \mathbb{F}_q[x]/\langle x^s-\alpha \rangle $ such that $g(x,\omega^{1+kr})=p_k(x)p_k''(x). $
From, equation (\ref{eq3}), we get that $g(x,y)\eta_{k}(y)=\eta_{k}(y) p_k(x)p_k''(x)$. \vspace{2mm}

\noindent Now, $g(x,y)=\sum_{k=0}^{\ell-1} g(x,y)\eta_{k}(y)=\sum_{k=0}^{\ell-1} \eta_{k}(y) p_k(x)p_k''(x)$ implies that $$g(x,y) \in \big\langle \eta_0(y) p_0(x) , \eta_{1}(y) p_1(x), \cdots , \eta_{\ell-1}(y) p_{\ell-1}(x) \big\rangle $$
Therefore, $$\mathcal{C} \subseteq \big\langle \eta_0(y) p_0(x) , \eta_{1}(y) p_1(x), \cdots , \eta_{\ell-1}(y) p_{\ell-1}(x) \big\rangle. $$
Also,  $I_k=\langle p_k(x) \rangle $ implies that $\eta_{k}(y)p_k(x) \in \mathcal{C}$ for all $k=0,1,\cdots,\ell-1$. Thus,
$$\mathcal{C}= \big\langle \eta_0(y) p_0(x) , \eta_{1}(y) p_1(x), \cdots , \eta_{\ell-1}(y) p_{\ell-1}(x) \big\rangle. ~~~~~~~~~~~~~~~~~~~~~~~~~~~~~~~~ \Box$$

\begin{theorem} \label{th2}
	Let deg $p_i(x)=a_i$ for $i=0,1,\cdots,\ell-1$, then a generator matrix of $\mathcal{C}$ is $$ G=\begin{pmatrix}
	p_0(x) \eta_0(y)\\
	xp_0(x) \eta_0(y)\\
	\vdots\\
	x^{s-a_0-1}p_0(x) \eta_0(y)\\
	p_1(x) \eta_1(y)\\
	xp_1(x) \eta_1(y)\\
	\vdots\\
	x^{s-a_1-1}p_1(x) \eta_1(y)\\
	\vdots\\
	p_{\ell-1}(x) \eta_{\ell-1}(y)\\
	xp_{\ell-1}(x) \eta_{\ell-1}(y)\\
	\vdots\\
	x^{s-a_{\ell-1}-1}p_{\ell-1}(x) \eta_{\ell-1}(y)
	\end{pmatrix}.	$$
\end{theorem}

\noindent {\bf Proof :}
It is enough to prove that rows of G are linearly independent. Suppose, if possible, there exist polynomials $m_0(x), m_1(x), \cdots ,m_{\ell-1}(x)$ in $\mathbb{F}_q[x]/\langle x^s-\alpha \rangle $, with deg $m_i(x) \leq s-a_i-1$, for ${i=0,1,2, \cdots ,\ell-1}$, such that $$m_0(x)\eta_0(y)p_0(x) + m_1(x)\eta_{1}(y)p_1(x) + \cdots +m_{\ell-1}(x)\eta_{\ell-1}(y)p_{\ell-1}(x)=0$$ in $\mathbb{F}_q[x,y]/\langle x^s-\alpha,y^\ell-\beta \rangle.$ Therefore, there exist polynomials $a(x,y) , b(x,y) \in \mathbb{F}_q[x,y]$ such that
\begin{equation} \label{eq4}
\sum_{j=0}^{\ell-1}m_j(x)\eta_{j}(y)p_j(x)=a(x,y)(x^s-\alpha) + b(x,y)(y^\ell-\beta).
\end{equation}

 \noindent Substituting $y=\omega^{1+kr} $ in equation (\ref{eq4}), we get
 $m_k(x)p_k(x)=a(x,\omega^{1+kr})(x^s-\alpha)$ because $\eta_{k}(\omega^{1+kr})=1$, $\eta_{j}(\omega^{1+kr})=0$ for $j \neq k$ and $(\omega^{1+kr})^\ell=\beta$. Since deg$( m_k(x)p_k(x)) \leq s-a_k-1+a_k < s$ , comparing coefficients of $x^s,x^{s+1},\cdots$ on both sides we find that  coefficients  of $a(x,\omega^{1+kr})$ are all zero and hence $a(x,\omega^{1+kr})=0$ $i.e.$ $m_k(x)p_k(x)=0$ in $\mathbb{F}_q[x]$. Therefore, $m_k(x)=0$ for all $k=0,1,\cdots,\ell-1$; and thus the rows of G form a generator matrix of $\mathcal{C}$. \hfill $\square$

\begin{cor}\normalfont The dimension of a two-dimensional $(\alpha,\beta )$-constacyclic code  $\mathcal{C}$ is given by $s\ell-a_0-a_1-\cdots-a_{\ell-1}$.\end{cor}

\subsection{Generator matrix of dual code}

We assume here that $\alpha, \beta \in \{1,-1\}$ so that  the dual code $\mathcal{C}^{\perp}$ of a two-dimensional $(\alpha,\beta )$-constacyclic code $\mathcal{C}$ is also an ideal  in $\mathbb{F}_q[x,y]/\langle x^s-\alpha,y^\ell-\beta \rangle.$ \vspace{2mm}

\noindent  As $ \dim(\mathcal{C}) + \dim(\mathcal{C}^{\perp})=s \ell$, therefore $\dim(\mathcal{C}^\perp)=a_0+a_1+\cdots+a_{\ell-1}$. As $p_j(x)$ are divisors of $x^s-\alpha$, there exist polynomials $p_j'(x) \in \mathbb{F}_q$ such that $p_j(x)p_j'(x)=x^s-\alpha$ for $j=0,1,\cdots,\ell-1$.  The following theorem gives the generators of the dual code $C^\perp$:
\begin{theorem} \label{th3}
Suppose that $\alpha,\beta \in \{1,-1\}$. Let $\mathcal{C}$ be a two-dimensional $(\alpha,\beta )$-constacyclic code of length $n=s.\ell$  as given in Theorem \ref{th1}, then the dual code

\begin{equation}\label{eq5}\mathcal{C}^\perp= \big\langle p_0'^*(x) \eta_0^*(y), p_1'^*(x) \eta_1^*(y),\cdots,p_{\ell-1}'^*(x) \eta_{\ell-1}^*(y)\big\rangle \end{equation}
and a generator matrix of $\mathcal{C}^\perp$ is given by,
$$ H=\begin{pmatrix}
	p_0'^*(x) \eta_0^*(y)\\
	xp_0'^*(x) \eta_0^*(y)\\
	\vdots\\
	x^{a_0-1}p_0'^*(x) \eta_0^*(y)\\
	p_1'^*(x) \eta_{1}^*(y)\\
	xp_1'^*(x) \eta_{1}^*(y)\\
	\vdots\\
	x^{a_1-1}p_1'^*(x) \eta_{1}^*(y)\\
	\vdots\\
	p_{\ell-1}'^*(x) \eta_{\ell-1}^*(y)\\
	xp_{\ell-1}'^*(x) \eta_{\ell-1}^*(y)\\
	\vdots\\
	x^{a_{\ell-1}}p_{\ell-1}'^*(x) \eta_{\ell-1}^*(y)
	\end{pmatrix}.	$$
\end{theorem}

\noindent {\bf Proof:} Let the code on right hand side of equation (\ref{eq5}) be denoted by $D$. To prove that $D \subset \mathcal{C}^\perp$ it is enough to prove, by Proposition 3, that
$p_j'^*(x) \eta_j^*(y)\in {\rm ann}(\mathcal{C})^\ast$ i.e, $p_j'(x) \eta_j(y)\in {\rm ann}(\mathcal{C})$ for each $j,~ 0\le j \le \ell-1$. Now for any $j,~  0\le j \le \ell-1$,  $$p_j'(x) \eta_j(y) p_k(x) \eta_k(y)=0 $$ for all $k, ~0\le k \le \ell-1$,  because when $k \neq j$, we have $\eta_j(y)\eta_k(y)=0$ and when $k=j$ we have $p_j'(x) p_j(x)=x^s-1=0$ in the ring $\mathcal{R}$. Hence $D \subset \mathcal{C}^\perp$. \vspace{2mm}

To prove that   $\dim(D)=a_0+a_1+\cdots+a_{\ell-1}=\dim (\mathcal{C}^\perp)$, we need to show that  the rows of $H$  are linearly independent. Suppose, if possible, there exist polynomials $m_0(x), m_1(x), \cdots ,m_{\ell-1}(x)$ in $\mathbb{F}_q[x]/\langle x^s-\alpha \rangle $, with $\deg m_i(x) \leq a_i-1$ for ${i=0,1,2, \cdots ,\ell-1}$ such that $$m_0(x)p_0'^*(x) \eta_0^*(y) + m_1(x)p_1'^*(x) \eta_0^*(y) + \cdots +m_{\ell-1}(x)p_{\ell-1}'^*(x) \eta_{\ell-1}^*(y)=0$$ in  $\mathbb{F}_q[x,y]/\langle x^s-\alpha,y^\ell-\beta \rangle.$ Therefore, there exist polynomials $a(x,y) , b(x,y) \in \mathbb{F}_q[x,y]$ such that
\begin{equation} \label{eq6}
\sum_{k=0}^{\ell-1}m_k(x)p_k'^*(x) \eta_k^*(y)=a(x,y)(x^s-\alpha) + b(x,y)(y^\ell-\beta).
\end{equation}

\noindent Let first $r=1$. Then by Lemma 5, $\eta_k^*(y)=b_k \eta_{\ell-2-k}(y)$, for some non-zero constant $b_k$. Also $\eta_{\ell-2-k}(\omega^{\ell-1-k})=\zeta_{\ell-1-k}(\omega^{\ell-1-k})=1$ and $\zeta_{\ell-1-j}(\omega^{\ell-1-k})=0$ for $j \neq k$.\vspace{2mm}

 \noindent Substituting $y=\omega^{\ell-1-k} $ in equation (\ref{eq6}) , we get
 $$b_k m_k(x)p_k'^*(x)=a(x,\omega^{\ell-1-k})(x^s-\alpha).$$  Since $\deg( m_k(x)p_k'^*(x)) \leq a_k-1+s-a_k < s$, comparing coefficients of $x^s,x^{s+1},\cdots$ on both sides we find that  coefficients  of $a(x,\omega^{\ell-1-k})$ are all zero and hence $a(x,\omega^{\ell-1-k})=0$ i.e.  $b_k m_k(x)p_k'^*(x)=0$ in $\mathbb{F}_q[x]$. Therefore, $m_k(x)=0$ for all $k=0,1,\cdots,\ell-1$. \vspace{2mm}

\noindent Let now $r=2$. Then by Lemma 5, $\eta_k^*(y)=b_k \eta_{\ell-1-k}(y)$. Also $\eta_{\ell-1-k}(\omega^{1+2(\ell-1-k)})$ $=\zeta_{1+2(\ell-1-k)}(\omega^{1+2(\ell-1-k)})=1$ and $\eta_{\ell-1-j}(\omega^{1+2(\ell-1-k)})=0$ for $j \neq k$.
Working as above we find that $m_k(x)=0$ for all $k=0,1,\cdots,\ell-1$. \vspace{2mm}

\noindent Therefore $D=\mathcal{C}^\perp$ and the rows of $H$ form a generator matrix of $\mathcal{C}^\perp$.
 \hfill $\square$\vspace{2mm}

\noindent{\bf Remark} If $\deg(p_j(x))=a_j=0$ for some $j,~0\le j \le \ell-1$, i.e. $p_j(x)=\lambda$, a constant, then the set $\{p_j'^*(x) \eta_{j}^*(y),~
	xp_j'^*(x) \eta_{j}^*(y),\cdots,
	x^{a_j-1}p_j'^*(x) \eta_{j}^*(y)\}$ is empty and so it does not contribute any rows in $H$.
 \subsection{Self-dual codes}
\begin{theorem}\label{th4} Suppose that $\alpha,\beta \in \{1,-1\}$. Then  a two-dimensional $(\alpha,\beta )$-constacyclic code $\mathcal{C}$ of length $n=s.\ell$  is self-dual if and only if
\begin{enumerate}[$\rm(i)$] \item $s\ell=2(a_0+a_1+\cdots+a_{\ell-1})$ \item   for every $k, 0\le k \le \ell-1$,
\begin{equation}\label{eq7}\begin{array}{lll} p_k'^*(x)= t_k(x)p_{\ell-2-k}(x),~ p_k(x)= t'_k(x)p'^*_{\ell-2-k}(x) & {\rm if} & \beta=1\\ p_k'^*(x)= t_k(x)p_{\ell-1-k}(x),~ p_k(x)= t'_k(x)p'^*_{\ell-1-k}(x) & {\rm if} & \beta=-1\end{array}\end{equation}
for some non-zero polynomials  $t_k(x)$, $t'_k(x)$ in $ \mathbb{F}_q[x]/\langle x^s-\alpha \rangle $.\end{enumerate}
\end{theorem}

\noindent {\bf Proof:}  Suppose $s\ell=2(a_0+a_1+\cdots+a_{\ell-1})$. By Theorem 3, $\mathcal{C}^\perp= \big\langle p_0'^*(x) \eta_0^*(y), p_1'^*(x) \eta_1^*(y),$ $\cdots,p_{\ell-1}'^*(x) \eta_{\ell-1}^*(y)\big\rangle. $ Therefore $\mathcal{C}^\perp \subset \mathcal{C}$ if  each  $p_k'^*(x)\eta_k^*(y)$ is a linear combination of rows of generator matrix $G$ of $\mathcal{C}$ as given in Theorem \ref{th2}. This is so if and only if there exist polynomials $u_j(x)$ in $ \mathbb{F}_q[x]/\langle x^s-\alpha \rangle $ such that
\begin{equation}\label{eq8} p_k'^*(x)\eta_k^*(y)=\sum_{j=0}^{\ell-1}u_j(x)p_j(x) \eta_j(y).\end{equation}

\noindent Again by Theorem 2, $\mathcal{C}= \big\langle p_0(x) \eta_0(y), p_1(x) \eta_1(y),\cdots,p_{\ell-1}(x) \eta_{\ell-1}(y)\big\rangle $, therefore $\mathcal{C} \subset\mathcal{C}^\perp$ if  each  $p_k(x)\eta_k(y)$ is a linear combination of rows of generator matrix $H$ of $\mathcal{C}^\perp$ as given in Theorem \ref{th3}. This is so if and only if there exist polynomials $v_j(x)$ in $ \mathbb{F}_q[x]/\langle x^s-\alpha \rangle $ such that
\begin{equation}\label{eq9} p_k(x)\eta_k(y)=\sum_{j=0}^{\ell-1}v_j(x)p'^*_j(x) \eta^*_j(y).\end{equation}

\noindent When $\beta=1$, we have $r=1$ and $\eta_k^*(y)=b_k \eta_{\ell-2-k}(y)$, by Lemma 5. Multiplying both sides of (\ref{eq8}) by $\eta_{\ell-2-k}(y)$ and  noting that $\eta_j(y)$ are primitive central idempotents,  $\mathcal{C}^\perp \subset \mathcal{C}$, if and only if
$$ p_k'^*(x)b_k \eta_{\ell-2-k}(y)=\sum_{j=0}^{\ell-1}u_j(x)p_j(x) \eta_j(y)\eta_{\ell-2-k}(y)=u_{\ell-2-k}(x)p_{\ell-2-k}(x) \eta_{\ell-2-k}(y)$$
i.e. if and only if
$$p_k'^*(x)= t_k(x)p_{\ell-2-k}(x)$$
for some polynomial $t_k(x)$.\vspace{2mm}

\noindent Further, (\ref{eq9}) can be rewritten as
\begin{equation}\label{eq10} p_k(x)\eta_k(y)=\sum_{j=0}^{\ell-1}b_jv_j(x)p'^*_j(x) \eta_{\ell-2-j}(y)= \sum_{i=0}^{\ell-1}b_{\ell-2-i}v_{\ell-2-i}(x)p'^*_{\ell-2-i}(x) \eta_{i}(y).\end{equation}
Multiplying both sides of (\ref{eq10}) by $\eta_{k}(y)$,  $\mathcal{C} \subset \mathcal{C}^\perp$, if and only if

$$ p_k(x) \eta_{k}(y)=b_{\ell-2-k}v_{\ell-2-k}(x)p'^*_{\ell-2-k}(x) \eta_{k}(y)$$
i.e. if and only if
$$p_k(x)= t'_k(x)p'^*_{\ell-2-k}(x)$$
for some polynomial $t'_k(x)$.\vspace{2mm}

\noindent When $\beta=-1$, we have $r=2$, $\eta_k^*(y)=b_k \eta_{\ell-1-k}(y)$ and the proof is similar.\vspace{2mm}

\begin{theorem}\label{th5} If $\beta=1$ and $\alpha=\pm 1$ then  a two-dimensional $(\alpha,1 )$-constacyclic code $\mathcal{C}$ of length $n=s.\ell$  can not be self-dual   if $\gcd(s,q)= 1$ assuming that  $s$ is odd if $\alpha=-1$. \end{theorem}

\noindent {\bf Proof : }   Suppose a two-dimensional $(\alpha,1 )$-constacyclic code is self-dual. On taking $k=\ell-1$ in equation (\ref{eq7}) we get
\begin{equation}\label{eq11} p_{\ell-1}'^*(x)= t_{\ell-1}(x)p_{\ell-1}(x) {\rm ~~ and~~} p_{\ell-1}(x)= t'_{\ell-1}(x)p'^*_{\ell-1}(x). \end{equation}

\noindent We have $x^s-\alpha=p_{\ell-1}(x)p'_{\ell-1}(x)$. If $\gcd(s,q)= 1$,  $x-\alpha$  divides exactly one of $p_{\ell-1}(x)$ and $p'_{\ell-1}(x)$ and not both. (If $\alpha =-1$, we assume that $s$ is odd.) The reciprocal of $x-\alpha$ is $\pm(x-\alpha)$.\vspace{2mm}

\noindent If $x-\alpha | p_{\ell-1}(x)$, from equation (\ref{eq11}) we have $x-\alpha | p_{\ell-1}'^*(x)$. This implies
$(x-\alpha)^* | (p_{\ell-1}'^*(x))^*$ i.e. $x-\alpha | p'_{\ell-1}(x)$ as $(f^*)^*=f$. This is not possible, when $\gcd(s,q)= 1$.  \vspace{1mm}

\noindent If $x-\alpha | p_{\ell-1}'(x)$,  we have $(x-\alpha)^* | p_{\ell-1}'^*(x)$ i.e. $(x-\alpha) | p_{\ell-1}'^*(x)$. This implies, from  (\ref{eq11}),
 $x-\alpha | p_{\ell-1}(x)$, again not possible.

\section{Examples}
Minimum distances of all the codes in following examples have been calculated by software MAGMA.
\begin{eg}
	\normalfont	Let $q=11$,  $\alpha =1$ , $\beta =-1$, $s=2$ and $\ell =5$. One finds that $\omega=2 $ is a $10^{th}$ root of unity in $\mathbb{F}_{11}^\ast$ such that $\omega ^{5} = -1$. Therefore
	\begin{equation*}
	\begin{array}{ll}
	y^5+1 &=(y-2)(y-2^3)(y-2^5)(y-2^7)(y-2^9) \vspace{2mm}\\
	&=(y+9)(y+3)(y+1)(y+4)(y+5)
	\end{array}
	\end{equation*}
	
	\noindent Thus,
	\begin{equation*}
	\begin{array}{ll}
	\eta_{0}(y) &=4y^4+8y^3+5y^2+10y+9 \vspace{2mm}\\
	\eta_{1}(y) &=5y^4+7y^3+y^2+8y+9 \vspace{2mm}\\
	\eta_{2}(y) &=9y^4+2y^3+9y^2+2y+9 \vspace{2mm}\\
	\eta_{3}(y) &=3y^4+10y^3+4y^2+6y+9 \vspace{2mm}\\
	\eta_{4}(y) &=y^4+6y^3+3y^2+7y+9 \vspace{2mm}\\
	\end{array}
	\end{equation*}
	
	\noindent We have $x^2-1=(x-1)(x+1)$. Suppose, $p_0(x)=p_4(x)=x+1$ and $p_1(x)=p_2(x)=p_3(x)=x-1$, then by Theorems \ref{th2} and \ref{th3},  generator matrices  of two dimensional $ \left( 1,-1 \right) $-constacyclic code $\mathcal{C}$ and $\mathcal{C}^\perp$ are given by
	$$
	G=\begin{pmatrix}
	(x+1)(4y^4+8y^3+5y^2+10y+9)\\
	(x-1)(5y^4+7y^3+y^2+8y+9)\\
	(x-1)(9y^4+2y^3+9y^2+2y+9)\\
	(x-1)(3y^4+10y^3+4y^2+6y+9)\\
	(x+1)(y^4+6y^3+3y^2+7y+9)
	\end{pmatrix},H=\begin{pmatrix}
	(1-x)(9y^4+10y^3+5y^2+8y+4)\\
	(x+1)(9y^4+8y^3+y^2+7y+5)\\
	(x+1)(9y^4+2y^3+9y^2+2y+9)\\
	(x+1)(9y^4+6y^3+4y^2+10y+3)\\
	(1-x)(9y^4+7y^3+3y^2+6y+1)
	\end{pmatrix}
	$$
\noindent respectively. The code $\mathcal{C}$ is not  self-dual as condition (ii) of Theorem 4 is not satisfied for $k=\ell-1=4$, i.e. $p_4'^*(x)=1-x \neq m(x) p_0(x)$.\vspace{2mm}

\noindent The codewords corresponding to rows of the matrix $G$ are \vspace{2mm}

\noindent $c_0=\begin{pmatrix}
	9&10&5&8&4 \\
	9&10&5&8&4
	\end{pmatrix}
	$,
	$c_1=\begin{pmatrix}
	-9&-8&-1&-7&-5\\
	9&8&1&7&5
	\end{pmatrix}
	$,
	$c_2=\begin{pmatrix}
	-9&-2&-9&-2&-9\\
	9&2&9&2&9
	\end{pmatrix}
	$,
	$c_3=\begin{pmatrix}
	-9&-6&-4&-10&-3\\
	9&6&4&10&3
	\end{pmatrix}
	$,
	$c_4=\begin{pmatrix}
	9&7&3&6&1\\
	9&7&3&6&1
	\end{pmatrix}.$\vspace{2mm}

\noindent The corresponding code $\mathcal{C}_1$ (see equation (\ref{eq1})) has a generator matrix
	$$G_1=\begin{pmatrix}
	&9&10&5&8&4&9&10&5&8&4&\\
	&-9&-8&-1&-7&-5&9&8&1&7&5&\\
	&-9&-2&-9&-2&-9&9&2&9&2&9&\\
	&-9&-6&-4&-10&-3&9&6&4&10&3&\\
	&9&7&3&6&1&9&7&3&6&1&
	\end{pmatrix}.
	$$
	It is a quasi-cyclic $[10,5,6]$  MDS code over $\mathbb{F}_{11}[x]$ . The dual $\mathcal{C}_1^\perp$ is also $[10,5,6]$  MDS code. The code $\mathcal{C}_1$ is  isodual but not  self-dual. \vspace{2mm}

\noindent The corresponding code $\mathcal{C}_2$ (see equation (\ref{eq2})) is a $(-1)$-quasi-twisted, isodual, $[10,5,6]$  MDS code over $\mathbb{F}_{11}[x]$.

\end{eg}
\begin{eg}\normalfont If in the above example we take $p_0(x)=p_4(x)=x+1$, $p_1(x)=p_3(x)=x-1$ and $p_2(x)=1$;  generator  matrices of $\mathcal{C}$ and $\mathcal{C}^\perp$ are
 $$G=\begin{pmatrix}
	(x+1)(4y^4+8y^3+5y^2+10y+9)\\
	(x-1)(5y^4+7y^3+y^2+8y+9)\\
	(9y^4+2y^3+9y^2+2y+9)\\
     x(9y^4+2y^3+9y^2+2y+9)\\
	(x-1)(3y^4+10y^3+4y^2+6y+9)\\
	(x+1)(y^4+6y^3+3y^2+7y+9)
	\end{pmatrix},H=\begin{pmatrix}
	(1-x)(9y^4+10y^3+5y^2+8y+4)\\
	(x+1)(9y^4+8y^3+y^2+7y+5)\\
	(x+1)(9y^4+6y^3+4y^2+10y+3)\\
	(1-x)(9y^4+7y^3+3y^2+6y+1)\end{pmatrix}
	$$
respectively. The corresponding codes $\mathcal{C}_1$  and $\mathcal{C}_2$ are $[10,6,5]$  MDS code over $\mathbb{F}_{11}[x]$ . The dual $\mathcal{C}_1^\perp$ and $\mathcal{C}_2^\perp$ are $[10,4,7]$  MDS code.

\noindent  $\mathcal{C}_1$ is quasi cyclic code of index 5 and  $\mathcal{C}_2$ is  $(-1)$-quasi-twisted of index 2.
\end{eg}
\begin{eg}
	\normalfont	Take $q=7$, $\alpha =-1$ , $\beta =2$, $s=3$ and $\ell =2$, so that order $r$ of $\beta$ is $3$.   One finds that $\omega=3 $ is a $r\ell^{th}$ i.e. $6^{th}$  root of unity in $\mathbb{F}_{7}^\ast$ such that $\omega ^{2} = 2$. Therefore,
	
	\begin{equation*}
	\begin{array}{l}
	y^2-2 =(y-3)(y-3^{4})
	=(y-3)(y-4)
	\end{array}
	\end{equation*}
	
	\noindent Thus,
	\begin{equation*}
	\begin{array}{l}
	\eta_{0}(y) =6y+4,~~
	\eta_{1}(y) =y+4
	\end{array}
	\end{equation*}
	
	\noindent As  $x^3+1=(x+1)(x^2-x+1)$, take $p_0(x)=x^2-x+1$ and $p_1(x)=x+1$, then by Theorem \ref{th2}, a generator matrix of two dimensional $ \left( -1,2 \right) $ -constacyclic code is given by
	$$
	G=\begin{pmatrix}
	(x^2-x+1)(6y+4)\\
	(x+1)(y+4)\\
	x(x+1)(y+4)\\
	\end{pmatrix}.
	$$
The corresponding codes $\mathcal{C}_1$  and $\mathcal{C}_2$ are $[6,3,4]$  MDS codes over $\mathbb{F}_{7}$.   $\mathcal{C}_1$ is $(-1)$-quasi-twisted  code of index 2 and  $\mathcal{C}_2$ is  $2$-quasi-twisted code of index 3.

\end{eg}

\begin{eg}
	\normalfont	Take $q=7$, $\alpha =-1=\beta$, $s=3=\ell$  so that $r=2$. One notes that $\omega =3$ is  $6^{th}$ root of unity in $\mathbb{F}_{7}^\ast$ satisfying $\omega ^{3} = -1$. Therefore,
	\begin{equation*}
	\begin{array}{l}
	y^3+1 =(y-3)(y-3^3)(y-3^5)
	=(y+4)(y+1)(y+2).
	\end{array}
	\end{equation*}
	
	\noindent Thus,
	\begin{equation*}
	\begin{array}{l}
	\eta_{0}(y) =6y^2+4y+5,~
	\eta_{1}(y) =5y^2-5y+5,~
	\eta_{2}(y) =3y^2+y+5 .
	\end{array}
	\end{equation*}
	
	\noindent In $\mathbb{F}_{7}[x]$, we have $x^3+1=(x+1)(x^2-x+1)$. Suppose, $p_0(x)=p_2(x)=x^2-x+1$ and $p_1(x)=x+1$, then by Theorem \ref{th2}, a generator matrix of two dimensional $ \left( -1,-1 \right) $-constacyclic code is given by
	$$
	G=\begin{pmatrix}
	(x^2-x+1)(6y^2+4y+5)\\
	(x+1)(5y^2-5y+5)\\
	x(x+1)(5y^2-5y+5)\\
	(x^2-x+1)(3y^2+y+5)
	\end{pmatrix}.
	$$
	The corresponding codes $\mathcal{C}_1$  and $\mathcal{C}_2$ are $(-1)$-quasi-twisted $[9,4,4]$, near MDS codes of index 3 over $\mathbb{F}_{7}$.  Their duals are  $[9,5,3]$, near MDS codes over $\mathbb{F}_7$.

\end{eg}

\begin{eg}
\normalfont Take $q=5$, $\alpha =1$, $\beta=-1$, $s=2=\ell$. One finds that $\omega=2 $ is $r\ell^{th}$ $i.e.$ $4^{th}$ root of unity in $\mathbb{F}_{5}^\ast$ satisfying $\omega ^{2} = -1$. Therefore,
\begin{equation*}
y^2+1 =(y-2)(y-2^3).
\end{equation*}

\noindent As $x^2-1=(x-1)(x+1)$, take  $p_0(x)=x-1$ and $p_1(x)=x+1$, so that $p_0'^\ast(x) =p_1(x) $ and $p_1'^\ast(x) =-p_0(x) $. Then by Theorem \ref{th2}, a generator matrix of two dimensional $ \left( 1,-1 \right) $ constacyclic code is given by
$$
G=\begin{pmatrix}
(x-1)(4y+3)\\
(x+1)(y+3)
\end{pmatrix}.
$$
The corresponding codes $\mathcal{C}_1$  and $\mathcal{C}_2$ are $[4,2,2]$  self-dual codes over $\mathbb{F}_{5}$.   $\mathcal{C}_1$ is quasi cyclic and   $\mathcal{C}_2$ is  $(-1)$-quasi-twisted code of index 2.

\end{eg}

\begin{eg}
	\normalfont Take $q=13$, $\alpha =1$, $\beta=-1$, $s=2$ and $\ell=6$. One finds that $\omega=2 $ is  $12^{th}$ root of unity in $\mathbb{F}_{13}^\ast$ satisfying $\omega ^{6} = -1$. Therefore,
	\begin{equation*}
	y^2+1 =(y-2)(y-2^3)(y-2^5)(y-2^7)(y-2^9)(y-2^{11}).
	\end{equation*}

	\noindent As $x^2-1=(x-1)(x+1)$. Take $p_0(x)=p_1(x)=p_2(x)=x-1$ and $p_3(x)=p_4(x)=p_5(x)=x+1$, then by Theorem \ref{th2}, a generator matrix of two dimensional $ \left( 1,-1 \right) $ constacyclic code is given by
	$$
	G=\begin{pmatrix}
	(x-1)(4y^5+8y^4+3y^3+6y^2+12y+11)\\
	(x-1)(3y^5+11y^4+10y^3+2y^2+3y+11)\\
	(x-1)(12y^5+7y^4+3y^3+5y^2+4y+11)\\
	(x+1)(9y^5+8y^4+10y^3+6y^2+y+11)\\
	(x+1)(10y^5+11y^4+3y^3+2y^2+10y+11)\\
	(x+1)(y^5+7y^4+10y^3+5y^2+9y+11)
	\end{pmatrix}.
	$$
The corresponding codes $\mathcal{C}_1$  and $\mathcal{C}_2$ are $[12,6,4]$  self-dual codes over $\mathbb{F}_{13}$.   $\mathcal{C}_1$ is quasi cyclic of index 6 and   $\mathcal{C}_2$ is  $(-1)$-quasi-twisted code of index 2.

\end{eg}

\end{document}